\documentstyle[12pt,a4]{article}
\begin{document}
\parskip=5pt plus 1pt minus 1pt

\begin{flushright}
{\bf LMU 08/97}\\
{\bf CERN-TH/97-200}
\end{flushright}

\vspace{0.2cm}

\begin{center}
{\Large\bf The Gluonic Decay of the $b$--Quark and the
$\eta '$--Meson\renewcommand{\thefootnote}{\fnsymbol{footnote}}
\footnote[1]{Supported in part by GIF contract I 0304--120.07/93 and
EEC contract CHRX--CT94--0579}}
\end{center}

\vspace{0.4cm}

\begin{center}
{\large\bf Harald Fritzsch}\\
\bigskip
\bigskip
{\it Theory Division, CERN, CH--1211 Geneva 23, Switzerland}\\
\bigskip
{\it and}\\
\bigskip
{\it Ludwig--Maximilians--Universit\"at M\"unchen, Sektion Physik,}\\
{\it Theresienstrasse 37, D--80333 M\"unchen}
\end{center}
\bigskip
\bigskip
{\bf \underline{Abstract:}}
The observed inclusive decay of $B$--mesons into $\eta ' + X$ is interpreted
as the consequence of the gluonic decay of the $b$--quark into an $s$--quark. As a
result of the QCD anomaly this decay proceeds partly as the decay
$b \rightarrow s + \eta '$, similar to $b \rightarrow s + J / \psi $. The
hadronic recoiling system is found to have a relatively large mass.
Analogously one expects a decay of the type $b \rightarrow s + \sigma $. The
branching ratios for these decays are large (of the order of 10 \%). The
results indicate that there is no room for an anomalously large
chromomagnetic decay mode of the $b$--quark. Gluon jets are expected to
exhibit an anomalously large tendency to fragment into $\eta '$- and
$\sigma$- mesons.\\
\vspace*{3cm}
\begin{flushleft}
{\bf LMU 08/97}\\
{\bf CERN-TH/97-200}
\end{flushleft}
\newpage
\noindent
Recently one has observed a relatively large rate for the decay of $B$--mesons
into an $\eta '$--meson and additional hadrons. One finds$^{1)}$:\\
\begin{equation}
Br \left( B \rightarrow \eta ' + X \right)
= (7.5 \pm 1.5 \pm 1.1) \times 10^{-4}
\end{equation}
under the constraint:\\
\begin{equation}
2.2 \,  GeV \le E (\eta ') \le 2.7 \, GeV.
\end{equation}
Furthermore the exclusive decay $B^{\pm } \rightarrow \eta ' K^{\pm}$ has been
observed$^{2)}$:
\begin{equation}
Br (B \rightarrow \eta ' K)
= \left( 7.8^{+2.7}_{-2.2} \pm 1.0 \right) \times 10^{-5}
\end{equation}
The latter is of the same order (albeit somewhat larger) as the decay
$B^{\pm} \rightarrow \pi ^{\pm}
K^0$, which is observed with a branching ratio of \quad
$2.3^{+1.1}_{-1.0} \cdot 10^{-5} $.\\
\\
The observation of an $\eta '$--meson in the final state of a $B$--decay is
of interest both with respect to the internal dynamics of such a decay
involving nonperturbative aspects of QCD and with respect to the validity of the standard
electroweak theory describing the decay.\\
\\
It is well known that the $\eta '$--meson plays a special role in the
dynamics of the strong interactions. Due to its QCD anomaly it evades to be a
massless Goldstone boson in the limit of the chiral $SU(3) \times SU(3)$
symmetry$^{3)}$. Due to its special character it is expected that the
$\eta '$--meson has an anomalously large coupling to gluonic field
configurations.$^{4)}$ Thus a possible connection between this feature, the QCD anomaly and the
$B$--decays involving the $\eta '$--meson, might exist. Such
possibilities have been discussed recently
in a number of papers$^{5, 6, 7, 8, 9)}$. In this note we shall discuss
the decay $B \rightarrow \eta ' + X$ from
a phenomenological point of view. It is shown that definite
conclusions can be drawn using simple and model-independent arguments.\\
\\
The decay $B \rightarrow \eta ' + X $ (either inclusive or exclusive) is a
decay, in which no charmed particles are present in the final state. Thus the
decay proceeds either via a charmless decay mode of the $b$--quark (see
below) or via the intermediate formation of a $\bar c c$--pair (via the decay
mode $b \rightarrow c (\bar c s)$), which annihilates to form the final
state containing an $\eta '$--meson. The most important decay channels would
be\\
a) $B \rightarrow J / \psi + X, \,  J / \psi \rightarrow \eta ' + X $
\qquad
b) $B \rightarrow \eta _c + X, \,  \eta _c \rightarrow \eta ' + X$.\\
\\
The possibility a) can easily be dismissed, since decays of the $J / \psi $
involving $\eta ' $ are seen of the level of $10^{-4}$ in branching ratio,
thus the corresponding $B$--decays are of the order of $10^{-6}$. The second
possibility was recently studied$^{5)}$, with the conclusion, that the expected
branching ratio for $B \rightarrow \eta ' + X $ could be at most $10^{-4}$,
i.\ e.\ small in comparison to the observed rate.\\
\\
It is well known that there is a relatively large mixing between the
$\eta '$--meson and the $\eta _c$--state$^{10)}$, which accounts for the decay
$J / \psi \rightarrow \eta ' \gamma $ (for an updated discussion, including 
the $\eta - \eta '$--mixing, see ref. (11)). Following ref. (10), we denote
the $\eta (\eta ')$--wave function as follows (in the absence of
a $\bar c c$--contribution):
\begin{equation}
\begin{array}{lll}
\eta & = & \frac{1}{\sqrt{2}} (\bar u u + \bar d d) \, {\rm cos} \,
\alpha - \bar s s \, {\rm sin } \, \alpha\\
\eta ' & = & \frac{1}{\sqrt{2}} (\bar u u + \bar d d) \, {\rm sin} \alpha \,
+ \bar s s \, {\rm cos} \, \alpha
\end{array}
\end{equation}
The mass spectrum of the pseudoscalar mesons as well as their radiative
decays indicate $\alpha \simeq 44^{\circ }$ (possible uncertainties
in this value will not be discussed here).
The large mixing between the various
$(\bar q q)$--contributions is due to the QCD anomaly. The latter is also
responsible or the $(\bar c c)$--admixture of the $\eta $ and $\eta '$--meson.
As discussed in ref.\ (10), the $\eta_c$--state can be written as:
\begin{equation}
\begin{array}{lll}
\eta_c & = & \bar c c + \varepsilon \cdot \eta + \varepsilon ' \cdot \eta '\\
\varepsilon & \cong & 0.010     \qquad \varepsilon ' \cong 0.024
\end{array}
\end{equation}
This gives a satisfactory description of the decay: $J / \psi \rightarrow
\eta (\eta ') + \gamma $. Thus the QCD anomaly which is responsible for
the large mixing in the pseudoscalar channel, accounts well for these
radiative decays. Likewise the decay $B \rightarrow \eta ' + X$ can proceed
via the $(\bar c c)$--admixture of the $\eta '$--meson, given by the mixing
parameter $\varepsilon '$, and one finds:
\begin{equation}
Br (B \rightarrow \eta ' X) = Br (B \rightarrow \eta _c X) \cdot
(\varepsilon ')^2 \preceq 10^{-2} \cdot (\varepsilon')^2
\approx 5 \cdot10^{-6}
\end{equation}
(Here we used $Br (B \rightarrow \eta _c X) \preceq 10^{-2} $
(see ref.(12)).
Thus the $\bar c c$--admixture of the $\eta '$--meson although anomalously
large, does not lead to a decay rate for $B \rightarrow \eta ' X$ at the
observed rate. Likewise one can see that the $\bar cc$--admixture does not
give a sizeable
contribution to the exclusive decay $B \rightarrow \eta ' K$. The decay
$B \rightarrow \eta _c K$ has not been observed, but its branching ratio is
expected to be smaller than for the decay $B \rightarrow J / \psi K$
(branching ratio of order $10^{-3}$). Thus one finds for the decay induced by mixing:
\begin{equation}
Br (B \rightarrow \eta ' K) \preceq 10^{-3} \cdot (\varepsilon ')^2 \simeq
5 \cdot 10^{-7}
\end{equation}
an effect far below the observed signal.
In general we conclude that the $\bar c c$-admixture of the $\eta
'$--meson
can be neglected for the discussion of the decays $B \rightarrow \eta ' X$ and
$B \rightarrow \eta ' K$.
Below we shall argue that the observed inclusive decay
$B \rightarrow \eta ' X$  is
indeed the first observation of the gluonic decay of the $b$--quark into an
$s$--quark.\\
\\
It is known that within the standard electroweak model the
$b$--quark can decay into an $s$-quark by emitting an on--shell or off--shell
gluon$^{13, 14)}$. The process proceeds via the transition of the $b$--quark into
a virtual $t$--quark and a $W$--boson, which combine to give an $s$--quark
under gluon--emission. If the gluon is on--shell, this process corresponds to
a chromomagnetic $b - s$ transition, described by the effective
Lagrangian\\
\begin{equation}
{\cal L}^{eff}_{b \rightarrow s} = - const. \frac{G_F}{\sqrt{2}} \cdot
\frac{g_s}{4 \pi ^2} \cdot m_b \, \bar b_R \sigma_{\mu \nu }
G^{\mu \nu} s_L \cdot V_{ts}
\end{equation}
\\
($G_{\mu \nu }$: gluonic field--strength, $V_{ts}: t-s$--transition element of the
weak current, the const. depends on details of perturbate QCD and is
estimated to be about 0.15).
If the gluon is off--shell, decaying afterwards into a $ \bar q q$--pair or
gluon--pair, the decay proceeds by a chromoelectric transition. Taking both
effects into account, one finds that about 1\% of all
$B$--decays should be of the type $b \rightarrow s +$ glue. The rate for this
process shall be denoted by $\Gamma ^g$. The low--order QCD--calculation
indicates that the chromomagnetic process is smaller than the
chromoelectric one.\\
\\
It is interesting to note that the gluonic decay of the $b$--quark would be
the dominant decay mode for the $b$--quark if the $c$--quark would be
heavier than the $b$--quark. In such a fictitious world the life--time of the
$B$--mesons would be about two orders of magnitude larger than in reality.
The question arises how the hadronic final state looks in the gluonic
$b$--decay. In the chromomagnetic decay the $b$--quark, essentially at rest
inside the $B$--meson, disintegrates into an almost massless $s$--quark
and a gluon. Both quanta are emitted back to back, each carrying a
momentum of about 2.3 GeV / c. (We shall use for our subsequent consideration
an effective $b$--mass of 4.6 GeV). This process is reminiscent of the decay
$b \rightarrow s + J / \psi $, discussed a long time ago$^{15)}$, except
in the
latter
process the energy of the emitted $s$--quark is about 1 GeV smaller than
in the process discussed here. The hadronic final state is formed by the
fragmentation of the $(s \bar q - g)$--system into hadrons.\\
\\
The chromoelectric decay proceeds in a similar way, except the emitted gluon
is not on--shell, i.\ e.\ the final system is a $(s \bar q - g^*)$--System.
Since the process is dominated by small $g^*$- masses, the fragmentation
process
should be similar, and we shall not discriminate between the two any longer.
In fact, in term of observable quantities a clear--cut distinction between the
chromomagnetic and chromoeletric reactions cannot be made.\\
The gluonic
$b$--decay is a reaction in which for the first time the conversion of a
single gluon of relatively low energy into hadrons can be studied
experimentally. A single color--octet gluon emitted by the decaying $b$--quark 
and leaving the hadronic debris of the previous $B$--meson with relativistic
speed can, of course, not be converted into a color--singlet system of
hadrons. However the color--octet nature of the gluon can easily be
``bleached'' by the interaction with one or several soft gluons
participating  in the process. If the mass of the
decaying $b$--quark were very high, say 20 GeV or more, one would observe a
gluon jet, identical to the gluon jets observed in other hadronic processes.
In our case, however, the ``gluon jet'' has only an energy of about 2.3 GeV.
A certain fraction of the decays will lead to glue--mesons, expected to have
a mass in the region above 1.6 GeV. These final states will be
complicated, involving a fairly large number of hadrons. We shall not
consider them any longer. An important fraction of the decays will involve
those hadrons which are known to have a strong affinity to the gluonic degrees
of freedom. In low energy hadron physics there are only two particles known
to have this property:
\begin{enumerate}
\item[a)] The $\eta '$--meson\\
The large $\eta '$--mass reflects the strong coupling of this meson to gluons
(see ref. 3, 4).
The matrix element $< 0 \mid A \mid \eta ' >$ of the gluonic operator
$A = \alpha _s \, G_{\mu, \nu } \,
\stackrel{\sim }{G^{\mu \nu }}$, which gives rise to the QCD--anomaly in the 
divergence  of the axial current bilinear in a particular quark field, e.\
g.\
$\bar u \gamma _{\mu }
\gamma _5 u$, is known to be large. The special properties of the
$\eta '$--meson due to its strong gluonic coupling as a consequence of 
the QCD-anomaly have been discussed a long
time ago$^{4)}$.

\item[b)] The $\sigma (f^0)$--meson\\
This $0^{++}$--meson is observed as a broad resonance: $M = (400 - 1200)$ MeV,
decaying predominantly into $ \pi \pi $. It is usually associated with the
$\sigma $--meson in chiral models of the $\pi $--nucleon--system. In QCD the
$\sigma $--state dominates the matrix elements of the trace of the
energy--momentum--tensor which is proportional to the scalar gluonic density
$G_{\mu \nu } \, G^{\mu \nu }$. Thus the transition element
$< 0 \mid \alpha_s \cdot G_{\mu } G _{\mu \nu} \, G ^{\mu \nu } \mid \sigma >$
is also anomalously large, and the $\sigma $--meson shows like the $\eta '$ a strong
affinity to gluons.\\
\end{enumerate}
The bleaching of the color-octet gluon emitted in the b-s-transition is
achieved by its interaction with one or several soft gluons. Finally a
color singlet gluonic system is emitted consisting of the original gluon
of relatively high energy, accompanied by at least one soft gluon. This
system can be described by a complicated sum of products of at least two
gluonic field strengths operators and their derivatives, corresponding to
the various spin and parity assignments of the multigluon system in
question. The hadronic matrix elements of these operators will finally
determine which types of hadrons are emitted in the decay. These states
will be primarily gluonic mesons, except for the special cases, in which
one deals with the pseudoscalar and scalar gluonic densities, which  at
low frequencies are strongly dominated by the $\eta '$
and $\sigma$--mesons. If the mass of the b-quark would be less than in
reality, say only 2.5 GeV, there would be no phase space for producing
gluonic mesons. Effectively all gluonic densities would be projected out, 
except for the scalar and pseudoscalar densities, due to the large
coupling to the $\sigma$ and $\eta '$--mesons. In this case the
gluonic decay
of the b-quark would lead exclusively to $\eta '$ or $\sigma$--mesons! \\
\\
If we restrict ourselves to the pseudoscalar and scalar densities, we can
write down an ansatz for the effective interaction between the quarks and the
gluonic densities as follows:\\
\begin{equation}
{\cal L}^{eff}  =  -const. \frac{G_F}{\sqrt{2}}V_{ts} 
\frac{\alpha_s}{4 \pi ^2} \cdot m_b \,  \bar b_R s_L \left( G_{\mu \nu}
G^{\mu \nu} + G_{\mu \nu } \stackrel{\sim} {G^{\mu \nu}}\right)
\end{equation}
where the const.\ absorbs the details of the non-perturbative aspects of
the
strong interactions involved in the decay process. No attempt is made here
to calculate this coefficient. Since at low frequecnies
the scalar and pseudoscalar densities are dominated by the corresponding
mesons, one concludes that a rather sizeable fraction
of the gluonic $b$--decay will be given by the decays $b \rightarrow s +
\eta '$ and $b \rightarrow s + \sigma $. The total gluonic decay rate is
given by
\begin{equation} 
\Gamma ^g = \Gamma ^{\eta '} + \Gamma ^{\sigma } + \stackrel{\sim }{\Gamma}
\end{equation}
where $\stackrel{\sim }{\Gamma} $ denotes in particular the contribution of
the large mass states (glue mesons etc.).\\
\\
It is interesting to study the kinematics of the decay $b \rightarrow s +
\eta '$. As an example we shall use a $b$--quark mass of 4.6 GeV and a
mass
of 0.2 GeV for the $s$--quark. For a $b$--quark at rest the energy of the
emitted $\eta '$--meson is 2.4 GeV, the momenta of both the $\eta '$ and the 
$s$--quark are 2.2 GeV/c. The invariant mass $M$ of the hadronic system
recoiling against the $\eta '$--meson can easily be calculated in the case in
which the $b$--quark is taken to be at rest in the rest system of the
$B$--meson. One finds M = 1.85 GeV.In reality the momentum of a $b$--quark 
inside the $B$--Meson varies between
zero and about 300 MeV. Thus the momenta of the emitted $\eta '$ should vary
around 2.4 GeV $(\pm \sim 300 $ MeV). Likewise the invariant mass of the
recoiling system should vary in the range 1.5 $\ldots $ 2.1 GeV. Both the    
momentum distribution of the $\eta '$--meson and the distribution of the
invariant mass should exhibit a Gaussean behaviour, reflecting  the
momentum distribution of the $b$--quark inside a $B$--meson. In fact,
the momentum distribution of the $\eta '$--meson can be interpreted as a 
measure of the momentum distribution of the $b$--quarks inside the
meson. \\
\\
The decay $b \rightarrow s +  \eta '$ differs in this respect from the decay
$b \rightarrow s + J / \psi $. There the momentum of the emitted $s$--quark
is much less, due to the large $\bar c c$--mass, and the invariant mass of
the
hadronic system recoiling against the $J / \psi $ is correspondingly much
lower such that a sizeable fraction of these decays are of the type
$b \rightarrow s K$ or $b \rightarrow s K^*$, in agreement with experiment.
In the case of the decays $b \rightarrow s + \eta '$ or $b \rightarrow s + 
\sigma $ one has little 
chance to find solely a $K$ or $K^*$--meson recoiling against the $\eta '$
or $\sigma $, due to the large average mass of the recoiling system. 
This agrees with the observed fact that the rate for
$B \rightarrow K \eta '$ is significantly smaller than the rate for
$B \rightarrow \eta ' X$.
The decays should be dominated by decays in which the recoiling hadronic
system has an invariant mass in the range between 1.5 and 2.1 GeV.\\ 
\\
The emitted $\eta '$--meson carries a sizeable energy (between about 2.1
GeV and 2.7 GeV). This energy region essentially coincides with the energy
cut made in the experimental analysis. It remains to be seen whether
the two main features discussed above (large invariant mass ($\sim $ 1.85
GeV)
of the recoiling system, large $\eta '$--energy ($\sim $ 2.4 GeV), narrow
momentum and mass distribution)  are
established in future experiments.\\
\\
The conversion of the gluon emitted in
the $b$--decay into an $\eta '$--meson is intrinsically a truly
non--perturbative effect, like the generation of the $\eta '$--mass. We
doubt whether it is useful to use perturbative methods, for example by
introducing an $\eta '- g - g$ vertex as suggested in refs.\ (5, 6). In
this 
approach  one would not expect that a single $\eta '$--meson is emitted,
while in our approach this is the case as a natural consequence of the
gluonic anomaly.\\
\\
The mechanism discussed in refs. (5, 6) leads to the decay $b \rightarrow
s \eta ' g$. Thus the elementary process is not a two--body decay. The
momentum distribution for the $\eta '$--meson is expected to be much broader
than in our case. While we do not expect that $\eta '$--mesons are found with
momenta less than 2.1 GeV and more than 2.7 GeV, in the $b \rightarrow
\eta '
g$--case this is expected. Likewise the invariant masses of the recoiling
systems should show a broad distribution. Thus the two possibilities
$b \rightarrow s \eta '$ and $b \rightarrow s \eta ' g$ can be distinguished
experimentally as soon as more detailed data become available.\\
\\
What has been said above for the decay $b \rightarrow s + \eta '$ can be
repeated for the decay $b \rightarrow s + \sigma $. Since the central value
of the $\sigma$--mass (800 MeV) is 160 MeV lower than the $\eta '$--mass, in
average the momentum of the emitted $s$--quark is slightly larger, but in
view of the large $\sigma $--width and the fluctuation of the $b$--momentum
inside the $B$--meson this effect can be neglected. Thus it is expected that
the $B$--meson decays into a $\sigma $--meson, such that the total energy is
in average about 2.4 GeV, recoiling against a hadronic system involving a
$K$--meson with an invariant mass just below 2 GeV.
The $\sigma $--meson decays into a $\pi \pi $--system. Due to the large
width of the $\sigma $--meson it is not easy to identify. Nevertheless the
simple two--body kinematics of the underlying decay
$b \rightarrow s + \sigma $ might help in the search for this decay.\\
\\
Finally we shall discuss the expected rate for the $\eta '$-- and
$\sigma $--decays of the $b$--quark, by comparing them to the total gluonic
decay rate, introducing the ratios
\begin{eqnarray}
r ( \eta ') & = & \Gamma (B \rightarrow \eta ' X) / \Gamma ^g\\
r ( \sigma ) & = & \Gamma (B \rightarrow \sigma X) / \Gamma ^g\nonumber
\end{eqnarray}
As argued above, these ratios will not be small. To get an order of
magnitude
estimate for $r(\eta ')$ and $r(\sigma )$, we consider the decay of the
$\eta _c$--meson. In lowest
order of QCD this meson decays into two gluons, each of them carrying an
energy of about 1.5 GeV (about 0.8 GeV less than the gluon emitted in the
gluonic $b$--decay). Applying the ideas discussed above to the
$\eta_c$--decay,
we would
expect that it decays relatively often into $\eta ' \eta ', \sigma \eta '$
and $\sigma \, \sigma $. Since the $\eta '$ is detected by its (small)
$\gamma \, \gamma $--decay mode, the $\eta ' \eta ' $--mode would be
difficult to identify, but the $\sigma \eta '$--mode, giving e.\ g.\ a final
state $\pi ^+ \pi ^- \eta '$, has indeed been identified as a leading
decay mode: $ Br \left( \eta _c \rightarrow \pi \pi \eta ' \right) =
4.1 \pm 1.7 \%$. The kinematics of the process is consistent with the
hypothesis that originally an $\eta ' -\sigma $--system is formed.
These observations support our idea that gluons at relatively low energy have
a sizeable
tendency to produce $\eta '$ or $\sigma $--mesons. We see no reason
why the fragmentation into an $\eta '$--meson should differ much from the
fragmentation into a $\sigma $--meson, and we expect that $r(\eta ')$ and
$r (\sigma )$ should be of similar order of magnitude.\\
\\
If we take $r (\sigma ) = r(\eta ')$, one finds $ r(\eta ') = r (\sigma )
\approx \left( Br (\eta_c \rightarrow \eta ' \sigma ) \right)^{1/2} \approx
0.2$. This estimate cannot be regarded as more than an estimate of order of
magnitude for $r (\eta ')$ and $ r(\sigma )$, in view of the large
uncertainties
involved, but it implies that e.\ g.\ $r (\eta ')$ is not small, but
surprisingly large,  of the order of 20\%, perhaps even larger, in
complete agreement with our expectation based on the dominance of the
gluonic densities by the corresponding isoscalar mesons. The contributions
of the exclusive
decays $b \rightarrow s + \eta '$ and $b \rightarrow s + \sigma$ to
$\Gamma ^g$ could make up a fairly large portion of all gluonic decays, say
20 $\ldots $ 50\%. \\
\\
For the inclusive decay $B \rightarrow \eta ' X$ we find
\begin{equation}
Br (B \rightarrow \eta ' X) = r (\eta ') \cdot \Gamma ^g / \Gamma ^{tot}
\end{equation}
As remarked earlier, the ratio $\Gamma ^g / \Gamma ^{tot}$ is expected in the
Standard Model to be about 1 \%. For $r (\eta ')$ = 10 \% we would obtain
$Br (B \rightarrow \eta ' + X) \approx 0.1\%$, not in disagreement with the
observed rate, in view of the uncertainty in the calculation of
$\Gamma ^g / \Gamma ^{tot}$. The branching ratio for the decay $B \rightarrow
\sigma + X$ should also be about 0.1\%.\\
\\
We add a remark concerning the decay $ B \rightarrow \eta + X$. If SU(3) were
an exact symmetry of the strong interactions, the decay $b \rightarrow s +
\eta $ would be forbidden. Due to SU(3) breaking effects it will be induced.
These breaking effects will show up in two different ways:
\begin{enumerate}
\item[a)] The $\eta $--meson is partly a SU(3) singlet and can
communicate via its singlet part with the pseudoscalar gluonic operator
$G_{\mu \nu} \stackrel{\sim }{G^{\mu \nu}}$. Thus the decay can proceed via
the singlet--octet--mixing in the wave function of the $\eta $--meson.
\item[b)] The gluonic coupling to $\bar u u / \bar d d$ and to $\bar s s$ in
the pseudoscalar channel violates SU(3)$^{4)}$. This effect also
influences the
decay rate for $b \rightarrow s + \eta$.
\end{enumerate}
 
We do not attempt a precise calculation of these effects, which in any case
would have a sizeable systematic error. However both effects play an
analogous role in the decays $J / \psi \rightarrow  \eta ' +
\gamma $,
which proceed via the gluonic mixing with the $\eta _c$--state. Using the
results of ref.\ (10), we find in terms of the parameters $\varepsilon $ and
$\varepsilon '$ (see eq. (4)):\\
\begin{equation}
\frac{\Gamma (b \rightarrow s + \eta)}{\Gamma (b \rightarrow s + \eta ')}
\approx \left( \frac{\varepsilon}{\varepsilon '} \right) ^2 \approx
\frac{1}{6}
\end{equation}\\
Phase space effects which are small due to the large $b$--mass have been
neglected in eq. (3). Thus we expect the rate for the inclusive decay
$B \rightarrow \eta + X$ to be about $10 ^{-4}$. It could hardly be more than
an order of magnitude less than the rate for the decay $B \rightarrow \eta '
+ X$ and should be seen eventually in the experiments. The momentum
distribution of the $\eta $ as well as the distribution of the invariant
masses of the recoiling systems are  identical to the $\eta '$--case, if
tiny  phase space corrections are neglected.\\
\\
In view of the fact that the semileptonic branching ratio for the
$B$--decays seems to be smaller than expected by theory and that there seems
to be a charm--deficit in the $B$--decays, it has been suggested that the
chromomagnetic decay $b \rightarrow s + g$ is abnormally large (branching
ratio 5 $\ldots $ 10\%), in disagreement with the Standard
Model$^{16, 17, 18)}$. (For recent calculations see refs. (18).) Our results
are in good agreement with the expectation within the Standard Model.
Therefore we conclude that there is no abnormally large chromomagnetic
$b$--decay. A large gluonic decay of the $b$--quark is likely not to
 be the reason for
the apparent charm deficit. Nevertheless  it could be that due to the
high proportion of exclusive decay modes on the level of quark decays
the estimate of the total gluonic decay rate based on perturbative QCD
is not correct. The actual rate could be enhanced by the nonperturbative 
effects discussed here, but due to duality arguments which provide a
link between perturbative estimates and resonance effects we doubt that
this effect could amount to more than a factor of two.  \\ \\
Our considerations support again the idea that due to the QCD--anomaly the
$\eta '$--meson plays an important and interesting dual role in hadronic
physics. It is a $\bar q q $--state, but shows like the $\sigma $--meson a
strong affinity to the gluonic sector of QCD. Gluons of relatively low energy
fragment with a large probability into an isolated $\eta '$--meson.
It would be interesting to search for an inclusive $\eta '$--signal in gluon
jets at high energies, e.g. at LEP. Our considerations suggest that high
energy
gluon jets should exhibit a leading $\eta '$--meson (see also 
refs. (4), (20) and (21)). Whether the strong $\eta$--signal,
seen in gluonic
jets at LEP by the L3-collaboration$^{22)}$ has anything to do with our
expectation remains to be seen. The observation of $\eta $ and
$\eta '$--mesons in the three-jet-events by the Aleph-collaboration$^{23}$
does not yet allow a firm conclusion, in view of the large experimental
uncertainties.\\
\\
We conclude: The observed inclusive decay $B \rightarrow \eta ' + X$ could
not be due to the primary $b$--quark decay $b \rightarrow $ c +  (cs). It is
the first signal for  the gluonic decay $b \rightarrow s +  glue $. 
About 10 $\ldots $ 20\% of these decays lead to the production of a single
$\eta '$--meson, which has a strong affinity to gluons due to the
QCD--anomaly. The decay can be viewed as the two--body decay $b \rightarrow
s + \eta ' $, analogous to $b \rightarrow s + J / \psi $. The hadronic system
recoiling against the $\eta '$ has a relatively large mass just below 2 GeV.
Likewise the decay $b \rightarrow s + \sigma $ should occur, which could be
identified by looking for decays $B \rightarrow \pi ^+ \pi ^- + X$, with
$E (\pi ^+ ) + E(\pi ^- ) \approx 2.3$ GeV and the invariant mass of the
$\pi \pi $--system being consistent with the $\sigma $--mass. 
Our results are in agreement with the prediction of the Standard Model
for the total gluonic decay rate of the $b$--quark. There is no or only
little room for an abnormally large chromomagnetic decay of the $b$--quark.\\
\\
{\bf Acknowledgements}:\\
I should like to thank M.\ Gronau, G.\ Hou, V. \ Khoze, P.\ Minkowski,
M.\ Neubert, U.\ Nierste, W.\ Schlatter, J.\ Steinberger, G.\ Veneziano and
G.\ Wolf for useful discussions.\\
\newpage
\noindent
\underline{References}\\
\begin{enumerate}

\item[1)] B.\ Behrens (CLEO coll.), talk presented at the conference on
$B$ Physics and $CP$ violation, Honolulu, Hawaii, March 1997

\item[2)] F.\ Wuerthwein  (CLEO coll.), talk presented at Rencontre de
Moriond, QCD and High Energy Hadronic Interactions, Les Arcs, March 1997

\item[3)] H.\ Fritzsch, M.\ Gell--Mann and H.\ Leutwyler, Phys.\ Lett.\
{\bf B47} (1973) 361.\\
	  E.\ Witten, Nucl.\ Phys.\ {\bf B149} (1979) 285\\
	  G.\ Veneziano, Nucl.\ Phys.\ {\bf B159} (1979) 213

\item[4)] H.\ Fritzsch and P.\ Minkowski, Nuovo Cimento 30 (1975) 393

\item[5)] D.\ Atwood and A.\ Soni, hep--ph/970 4357

\item[6)] W.\ S.\ Hou and B.\ Tseng, hep--ph/970 5304

\item[7)] I.\ Halperin and A.\ Zhitnitsky, hep--ph/9704412
	  
\item[8)] A.\ Ali and G.\ Greub, DESY--97--126, hep--phy/9707251\\
	   A.\ Dighe, M.\ Gronau and J.\ Rosner, CERN--TH/97--185,
	   EFI 97--34, hep--ph/9707521

\item[9)] A.\ L.\ Kagan and A.\ A.\ Petrov, hep--ph / 970 7354

\item[10)] H.\ Fritzsch and D.\ Jackson, Phys.\ Lett.\ {\bf 66B} (1977) 365

\item[11)] F.\ Gilman and R.\ Kauffman, Phys.\ Rev.\ 36D (1987) 2761

\item[12)] Particle Data Group, Phys.\ Rev.\ {\bf D54} (1996) 1

\item[13)] W.\ S.\ Hou, A.\ Soni and H.\ Steger, Phys.\ Rev.\ Lett.\
{\bf 59} (1987) 1521

\item[14)] W.\ S.\ Hou, Nucl.\ Phys.\ {\bf B308} (1988) 561

\item[15)] H.\ Fritzsch, Phys.\ Lett.\ {\bf 86B} (1979) 343\\
J.\ H.\ K\"uhn, S. Nussinov and R. R\"uckl, Z. Phys. {\bf C5} (1980) 117\\
J.\ H.\ K\"uhn and R. R\"uckl, Phys.\ Lett.\ {\bf 135B} (1984) 477

\item[16)] H.\ Fritzsch, in:  Results and Perspectives in Particle Physics,\\
M.\ Greco ed., Proceedings of the 1991 Rencontre de Physique, La Thuile,
Italy, p.\ 177

\item[17)] B.\ G.\ Grzadkowski and W.\ S.\ Hou, Phys.\ Lett.\ {\bf B272}
(1991) 383

\item[18)] A.\ L.\ Kagan, Phys.\ Rev.\ {\bf D51} (1995) 6196

\item[19)] A.\ Lenz, U.\ Nierste and G.\ Ostermaier, DESY 97--119,
hep--ph/9706501\\
M.\ Cinchini, E.\ Gabrielli and G.\ F.\ Guidice,
Phys.\ Lett.\ {\bf B388} (1996) 353.

\item[20)] I.\ Montvay, Phys.\ Lett. {\bf 84B} (1979) 331\\
C.\ Peterson and T.\ Walsh, Phys.\ Lett.\ {\bf 91B} (1980) 455\\
L.\ Azimov, Y.\ Dokshitser and V.\ Khoze, Sov.\ Phys.\ Uspekhi 23 (1980) 732

\item[21)] V. Khoze and W. Ochs, Int. J. of Mod. Phys.{\bf A12} (1997) 2949

\item[22)] M. Acciari et al., Phys.\ Lett.\ {\bf 311B} (1996) 129

\item[23)] Aleph--Coll., Ref.\ 598, submitted to the 1997 EPS--HEP--Conf.,
Jerusalem
\end{enumerate}
\end{document}